# Evaluating the Impact of a Specialized LLM on Physician Experience in Clinical Decision Support: A Comparison of Ask Avo and ChatGPT-4

Daniel Jung[1], Alex Butler[2], Joongheum Park[3], Yair Saperstein[3]

1. School of Medicine, University of Missouri, Kansas City, MO USA
2. Department of Pediatrics, Boston Children's Hospital, MA USA
3. Beth Israel Deaconess Medical Center, Harvard University, Cambridge, MA USA
4. Department of Medicine SUNY-Downstate-Health Science University, Brooklyn, NY USA

## Abstract

The use of Large language models (LLMs) to augment clinical decision support systems is a topic with rapidly growing interest, but current shortcomings such as hallucinations and lack of clear source citations make them unreliable for use in the clinical environment. This study evaluates Ask Avo, an LLM-derived software by AvoMD that incorporates a proprietary Language Model Augmented Retrieval (LMAR) system, in-built visual citation cues, and prompt engineering designed for interactions with physicians, against ChatGPT-4 in end-user experience for physicians in a simulated clinical scenario environment. Eight clinical questions derived from medical guideline documents in various specialties were prompted to both models by 62 study participants, with each response rated on trustworthiness, actionability, relevancy, comprehensiveness, and friendly format from 1 to 5. Ask Avo significantly outperformed ChatGPT-4 in all criteria: trustworthiness (4.52 vs. 3.34, $p<0.001$), actionability (4.41 vs. 3.19, $p<0.001$), relevancy (4.55 vs. 3.49, $p<0.001$), comprehensiveness (4.50 vs. 3.37, $p<0.001$), and friendly format (4.52 vs. 3.60, $p<0.001$). Our findings suggest that specialized LLMs designed with the needs of clinicians in mind can offer substantial improvements in user experience over general-purpose LLMs. Ask Avo's evidence-based approach tailored to clinician needs shows promise in the adoption of LLM-augmented clinical decision support software.

## Introduction

Large language models (LLMs) have garnered significant interest for their potential use in the clinical decision-support space ((1–3)). The ability of these models to process vast amounts of information and provide relevant insights quickly presents a transformative opportunity for practicing evidence-based medicine in healthcare. However, the current state of LLMs is known for frequent shortcomings. These shortcomings include hallucinations where responses sound plausible but contain incorrect or completely unrelated responses((4–6)). Another shortcoming is a lack of clarity on the sources cited to generate a response which provides a degree of uncertainty or untrustworthiness that makes adoption in the healthcare space challenging ((7)).

This study aims to evaluate whether Ask Avo, an LLM-derived software designed to retrieve information within a predefined set of guidelines using Language Model-Augmented Retrieval (LMAR), visual indications for direct citation, and system prompts tailored for physician interactions, can improve clinicians' experience in utilizing clinical decision support systems. Ask Avo is designed to provide precise and contextually relevant information by integrating directly cited sources and generating directly linked citations, enhancing trust and usability. Our primary goal is to determine whether Ask Avo can gain a better subjective user experience to physicians compared to general-purpose LLMs like OpenAI's GPT-4 in clinical decision support, addressing the critical issues of trust, actionability, relevance, comprehensiveness, and user-friendly format. By focusing on these aspects, we aim to explore the relationship of transparency and reliability in software design in the healthcare setting with trust and clinical utility as determined by a clinician.

## Methods

### Question Selection

Ten clinical guidelines were selected from a large corpus of consensus clinical guidelines shared by American Clinical Specialty Organizations. Each clinical guideline covered a different topic to avoid overlapping information. A clinical question was generated relevant to the condition or disease covered in a consensus guideline which could be answered with directly sourced information. The full list of clinical guideline documents is included below:

### LLM Preparation

Ask Avo is designed for use in environments where curated guideline documents are uploaded to the Ask Avo server. This setup allows Ask Avo to utilize LMAR tailored to each institution's specific needs, making it a highly adaptable tool for various clinical settings. For this study, the ten selected clinical guideline documents were uploaded to the Ask Avo server database, ensuring that Ask Avo could draw upon this curated information for its responses. This approach was intended to simulate Ask Avo's real-world use case, emphasizing its "out of the box" readiness without requiring additional setups such as document uploads or prompt engineering.

In contrast, ChatGPT-4 was not provided with these documents, reflecting its intended use case as a more general model. This difference highlights the practical advantages of Ask Avo's tailored approach that is designed to offer immediate utility in clinical settings without extensive configuration or customization.

### Participant Recruitment

Participants were recruited through social media posts and public outreach, specifically targeting physicians. A total of 62 participants completed the survey.

### Response Collection

A Google Form was used for survey responses. Following the clinical question in Ask Avo and GPT-4, participants were asked to grade the response from each model on a scale from 1 to 5, with 1 being the least and 5 being the most, in terms of the following five metrics: trustworthiness, actionability, relevancy, comprehensiveness, and readability or user-friendly format. Each metric had accompanying descriptions that further defined the metrics–trustworthiness: "How much do you trust the generated response?", actionability: "Would you feel comfortable using the generated response in a patient care scenario? Assume there are no legal or administrative barriers to using AI in your clinical care setting." relevancy: "Do you feel the generated response effectively understood and answered your question in a relevant way?", comprehensiveness: "Do you feel the generated response addressed all of the components of your question? In other words, you do not need to search for additional answers (outside of calculators or estimation tools).", and friendly format: "Did you feel the response was organized in a way that was easy to follow?" Each metric was also accompanied by an optional comment box for additional feedback, allowing participants to provide their insights into their experiences with each platform.

Participants were encouraged to correspond with both LLMs in a conversational thread until they were satisfied with the responses or deemed no further interaction was necessary. After interacting with the LLMs, participants completed the survey for each question, providing ratings and optional commentary on their experiences.

During the response collection phase, the study team identified an error in the Ask Avo system which was providing a response that did not correlate with provided guidelines. This error was triggered in only two clinical scenarios and was due to a logic error in the prompt. This bug was fixed mid-way through data collection so to avoid bias in the survey responses, these questions were removed from analysis for all participants, leaving eight clinical scenarios for analysis.

## Results

### Survey Scores

A paired sample T-test comparing ratings for Ask Avo and GPT4 was used to determine the significance of the results. Ask Avo significantly outperformed ChatGPT-4 in all measured criteria: Trustworthiness: 4.52 vs 3.34 (+35.30%, p<0.001), Actionability: 4.41 vs 3.19 (+38.25%, p<0.001), Relevancy: 4.55 vs 3.49 (+30.28%, p<0.001), Comprehensiveness: 4.50 vs. 3.37 (+33.41%, p<0.001), Friendly Format: 4.52 vs. 3.60 (+25.48%, p<0.001).

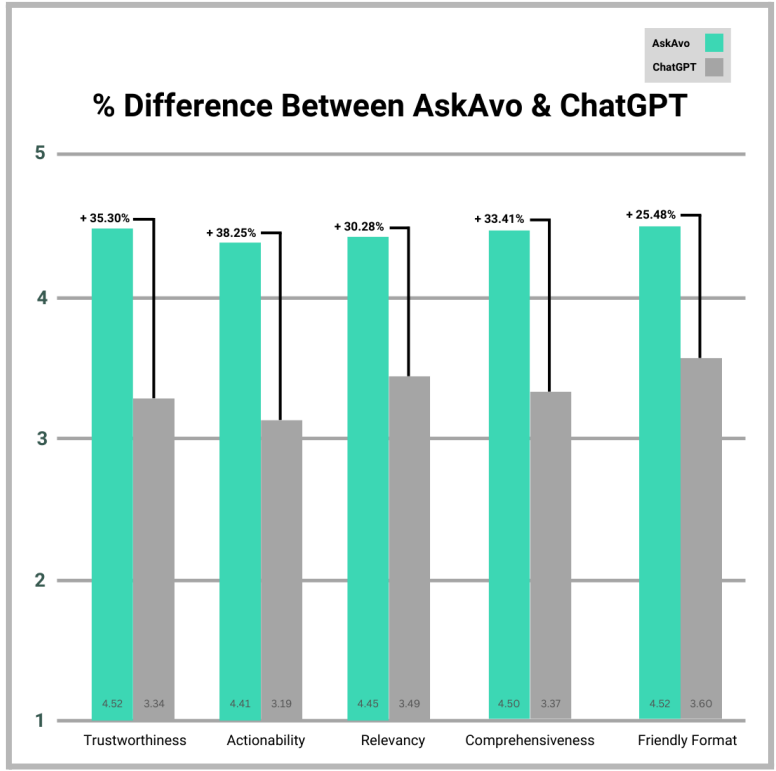

### Subjective Participant Feedback

Participants appreciated the direct citation feature, noting that the ability to easily visualize cited sources within the same interface significantly increased their trust and comfort with the information provided. One participant remarked, "I love the citations built directly into the source and the ability to easily pull those citations up on the same tab." The AI Fact-Check option, which clarifies the limitations of the information, was also positively received, with one user stating, "I love the 'AI Fact-Check' option that makes the limitations clear as well. This is excellent.."

Participants frequently mentioned Ask Avo's ability to provide specific, actionable advice as an advantage over ChatGPT. Many described responses from Ask Avo as more concise and focused, indicating a better understanding of the questions posed. One user noted, "Much more concise and focused than what was covered in GPT. Avo seemed to listen to the question." This focus on relevance and effectiveness was seen as a key strength, with another user highlighting, "This gives highly specific, actionable advice." The inclusion of information on treatment duration, effectiveness, and potential challenges faced by patients and care teams was also highly valued. A participant mentioned, "I appreciate that it includes duration, effectiveness, challenges that patients and the care team may face."

However, some criticisms were noted. Participants mentioned that Ask Avo's responses could be text-heavy and suggested that better organization into tables or sections could improve comprehension. One user suggested, "I think it is too text-heavy and organizing it into a table or something or having better headers, clear sections, and formatting would make it easier to comprehend." Some users found the responses lacking specific medication choices and dosing recommendations, which are critical for clinical decision-making: "Missing specific medication choices and dosing recommendations,". Hallucinations were also noted, with a participant commenting: "Mentions a group B and E that I don't know what that means." There were also comments about unclear steps and unfamiliar groupings, indicating that some responses might still benefit from refinement: "The steps here are very confusing."

Ask Avo, an LLM-derived software by AvoMD that incorporates a proprietary Language Model Augmented Retrieval (LMAR) system, in-built visual citation cues, and prompt engineering designed for interactions with physicians, against ChatGPT-4 in end-user experience for physicians in a simulated clinical scenario environment

## Discussion

These results indicate that Ask Avo provides a significantly better user experience across all evaluated criteria. The tailored design of Ask Avo, with its integration of direct citation cues, LMAR, and prompt engineering appears to find better acceptance by clinicians in settings of simulated clinical-decision support scenarios. This suggests that specialized LLMs, designed with specific clinical applications in mind, can offer substantial improvements in terms of trust, actionability, relevancy, and overall usability.

This study demonstrates that recognized shortcomings in current LLM platforms can be addressed with targeted solutions that enhance clinician perceptions when implemented. Most significantly, providing a curated amount of source material and generating in-line text citations significantly improve the trustworthiness of the responses to questions. Finally, individual participant feedback highlighted the utility of these changes as well as the ability to interrogate generated responses using the AI Fact-Check option. Taken together, this suggests that modifications to LLM platforms such as those found in Ask Avo can significantly improve the usability of LLM platforms for clinicians and may allow for their utility at the point of care in a clinical setting.

While specific shortcomings were addressed, Ask Avo was not without fault. Some users found its responses concise and comprehensive while others noted lengthy responses and incomplete information provided. This variety in responses suggests user-specific preferences and highlights that future improvements should address customization or adjustment at the user-level to further improve usability.

As LLMs continue to develop, their utility in the clinical field will also rise and there is currently a great amount of research into how this technology can best be deployed to help patients and providers((1–3)). A significant barrier to this adoption has been trustworthiness and the commonly cited 'black box problem' which stands as a significant impediment to adoption ((8)). Specific adjustments provided with the Ask Avo system demonstrate that these shortcomings can be addressed with significant subjective improvement while serving a critical need for providing accurate, actionable, and reliable information. This study emphasizes the future potential for specialized LLMs to transform clinical decision support while highlighting the next steps for further improvement.

## Limitation

This study did not directly assess the objective accuracy of the LLM responses as a metric, focusing instead on the subjective experiences of using an LLM designed for clinician use. Additionally, the study was conducted in a simulated scenario rather than real patient-physician encounters, which could influence the generalizability of the findings. Future studies should consider evaluating the accuracy and real-world applicability of these models in clinical settings.

## Conclusion

This study found that physicians experienced significantly better subjective outcomes using Ask Avo, which integrates LMAR and visible cues for direct citation, in a simulated clinical decision making environment. Ask Avo provided significantly greater trust, actionability, relevance, comprehensiveness, and user-friendly format compared to ChatGPT-4. These findings suggest that specialized LLMs designed with clinical needs in mind can offer substantial improvements over general-purpose LLMs, potentially transforming the landscape of digital health tools. As LLM technology continues to evolve, integrating specialized LLMs such as Ask Avo into clinical workflows could enhance evidence-based practice and improve patient care outcomes. However, it remains essential for the medical community to continue evaluating and refining these tools to ensure their accuracy, reliability, and overall effectiveness in real-world clinical settings. This study highlights the potential for tailored LLMs to significantly impact clinical decision support, advocating for further development and rigorous testing to fully realize their benefits in healthcare.